\documentstyle[12pt,epsfig]{article}
\newcommand{\simlt}  {\raisebox{-.6ex}{$\stackrel{\textstyle <}{\sim}$}}
\newcommand{\simgt}  {\raisebox{-.6ex}{$\stackrel{\textstyle >}{\sim}$}}
\begin{document}
\begin{flushright}
hep-ph/0202074 \\
RAL-TR-2002-002 \\
7 Feb 2002 \\
\end{flushright}
\begin{center}
{\Large
Tri-Bimaximal Mixing
and the Neutrino Oscillation Data.
}
\end{center}
\vspace{1mm}
\begin{center}
{P. F. Harrison\\
Physics Department, Queen Mary University of London,\\
Mile End Rd. London E1 4NS. UK \footnotemark[1]}
\end{center}
\begin{center}
{and}
\end{center}
\begin{center}
{D. H. Perkins\\
Nuclear Physics Laboratory, University of Oxford\\
Keble Road, Oxford OX1 3RH. UK \footnotemark[2]}
\end{center}
\begin{center}
{and}
\end{center}
\begin{center}
{W. G. Scott\\
Rutherford Appleton Laboratory\\
Chilton, Didcot, Oxon OX11 0QX. UK \footnotemark[3]}
\end{center}
\vspace{1mm}
\begin{abstract}
\baselineskip 0.6cm
\noindent
Following 
recent results from the
SNO solar neutrino experiment
and the K2K long-baseline neutrino experiment,
the combined existing data on neutrino oscillations
now point strongly to a specific form
for the lepton mixing matrix,
with effective bimaximal mixing
of $\nu_{\mu}$ and $\nu_{\tau}$
at the atmospheric scale
and effective trimaximal mixing
for $\nu_e$ with 
$\nu_{\mu}$ and $\nu_{\tau}$
at the solar scale
(hence `tri-bimaximal' mixing).
We give simple mass-matrices
leading to tri-bimaximal mixing,
and discuss its relation 
to the 
Fritzsch-Xing democratic ansatz. 
\end{abstract}
\begin{center}
{\em To be published in Physics Letters B}
\end{center}
\footnotetext[1]{E-mail:p.f.harrison@qmul.ac.uk}
\footnotetext[2]{E-mail:d.perkins1@physics.oxford.ac.uk}
\footnotetext[3]{E-mail:w.g.scott@rl.ac.uk}
\newpage
\baselineskip 0.6cm

\noindent {\bf 1. Motivation} \\

\noindent
The first  data
from the Sudbury
solar neutrino experiment (SNO) \cite{sno}
have dramatically confirmed the
long-standing HOMESTAKE \cite{home}
solar neutrino result
with regard to
the high-energy charged-current $\nu_e$-rate
(SNO/BP2001 $=0.347 \pm 0.029$,
HOME-STAKE/BP2001$=0.337 \pm 0.030$).
At the same time, 
the comparison with 
the existing rather precise SUPER-K (SK) result 
for solar-neutrino electron elastic scattering
(SK/BP2001 $=0.459 \pm 0.007$ \cite{sk1258})
(which includes a neutral-current contribution)
has provided a significant first
cross-check of the Bahcall-Pinsonneault
(BP2001) \cite{bp2001}
standard solar model 
calculation of the $^8$B flux
in these experiments,
so that it now
seems reasonable to conclude
that the suppression for $^8$B neutrinos
is probably close to $ \sim 1/3$.
In detail, 
the SNO, HOMESTAKE and SUPER-K
experiments have different thresholds
and response functions
(eg.\ HOMESTAKE is expected to
include a $\sim 15$\% 
contribution from $^7$Be neutrinos)
but such effects are readily 
taken into account \cite{fog1}.

By comparison,
the low energy charged-current $\nu_e$ rate
as sampled in the gallium-based experiments
SAGE \cite{sage}, GALLEX \cite{gall} and GNO \cite{gno} 
is known to be less suppressed
(SAGE/BP2001 $= 0.59 \pm 0.07$,
$\langle$GALLEX GNO$\rangle$/BP2001 $= 0.58 \pm 0.07$).
The gallium experiments
are 
sensitive to neutrinos from the pp chain 
and are only marginally 
affected by $^8$B neutrinos
(with expected signal contributions of $\sim 60\%$ pp,
$\sim 30\%$ $^7$Be and $\sim 10\%$ $^8$B
in the standard model \cite{bp2001}).
We have previously emphasised,
within the context of the 
original trimaximal mixing scenario \cite{hps1},
the consistency of the gallium suppression
with $5/9 \simeq 0.56$
(this consistency survives today
at the $1.2\, \sigma$ level, 
even allowing for the reduced
$^8$B contribution in gallium 
in the LMA solution, see below).

Thus
energy-dependence of the 
solar suppression 
is implicit,
with the latest general 
fits \cite{fog1} to the solar data
favouring the so-called
large-angle (LMA) MSW 
\cite{msw} 
solution.
The long-standing 
small-angle (SMA) MSW solution
is now essentially ruled out,
while the so-called LOW and VO solutions
are of marginal significance only 
\cite{fog1}.
The LMA solution 
is illustrated in Fig.~1
for several possible mixing angles.
In the LMA solution
the base of the MSW `bathtub'
is arranged to account for
the strong suppression at high-energy
where matter effects dominate. 
At lower energies
(for the same solar core density) 
the suppression reverts 
to its vacuum level 
outside the bathtub,
accounting 
for the gallium data
(the far high-energy end of the bathtub
plays no role in the LMA solution).
No significant day-night asymmetry
is observed 
\cite{sk1258} so that
the latest
LMA fits \cite{fog1} 
now prefer
a mass-squared difference
towards the higher end of the range
$\sim 10^{-5}-10^{-4}$ eV$^2$
(the curves of Fig.~1 are
for a representative
$\Delta m'^2 = 5 \times 10^{-5}$ eV$^2$).

Interestingly,
the trimaximal model \cite{hps1}
is known \cite{hps2}
to predict a `$5/9-1/3-5/9$' LMA bathtub
(see Fig.~3 of Ref.~\cite{hps2})
which could certainly be exploited to fit 
the current solar data in isolation.
However, within the trimaximal model,
the associated mass-squared difference
would then necessarily be 
the larger mass-squared difference
(compare Figs.~2 and 3 of Ref.~\cite{hps2})
and would thus be {\em in}consistent
with the data on atmospheric neutrinos,
which seem to
require a mass-squared difference
some two orders of magnitude greater
$\sim 10^{-3}-10^{-2}$ eV$^2$ 
\cite{atm}.

Indeed, the other
important new experimental input
motivating the present analysis,
is the currently emerging data
from the K2K 
long-baseline experiment \cite{k2k},
tending to confirm \cite{fog2}
the mass-squared difference
from the atmospheric neutrino experiments 
$\Delta m^2 \simeq 3 \times 10^{-3}$ eV$^2$,
clearly well above the solar
mass-squared difference
defined by the LMA solar fits
and, in particular, subject
to the CHOOZ \cite{chooz} and PALO-VERDE \cite{palov}
reactor limits on $\bar{\nu}_e$-mixing,
which imply $|U_{e3}|^2$ $\simlt$ $0.03$,
for $\Delta m^2$ $\simgt$ $10^{-3}$ eV$^2$.
($U$ denotes the lepton mixing or
MNS matrix, \cite{mns}).
Note that,
in this last respect,
the new K2K results
strongly disfavour the 
original trimaximal model.

An obvious solution
is to sacrifice the economy
of the original trimaximal model
and acknowledge
(in line with most other
phenomenological analyses)
the existence
of {\em two} distinct mass-squared difference scales,
$\Delta m^2 >> \Delta m'^2$,
controlling respectively the behaviour of
atmospheric and of solar neutrinos
Within this context,
the totality of the data 
clearly point
to a particuluar
form for the lepton mixing matrix,
which turns out
to be closely related
to the trimaximal hypothesis,
and which is given below. \\

\noindent {\bf 2. The Trend of the Data} \\

\noindent
The atmospheric neutrino results are known 
to point strongly to 
{\em two}fold maximal 
$\nu_{\mu} - \nu_{\tau}$ mixing
(or at least to effective
twofold maximal $\nu_{\mu}-\nu_{\tau}$ mixing
at the atmospheric scale).
In particular the SUPER-K zenith angle
distribution for multi-GeV
`$\mu$-like' events (Fig.~2a)
clearly indicates a suppression 
of upward $\nu_{\mu}$ 
with respect to downward $\nu_{\mu}$ 
by a factor of $\sim 1/2$
(from Fig.~2 the up-to-down ratio
for multiGeV muons is 
$(U/D)_{\mu} \simeq 0.53 \pm 0.05$
for zenith angles $| \cos \theta | > 0.20$).
By contrast, the corresponding distribution
for `e-like' events (Fig.~2b)
appears to be completely unaffected,
\mbox{$(U/D)_e \simeq 1.09 \pm 0.12$}.
In Fig.~2a
the solid curve is the full oscillation
curve for twofold maximal 
$\nu_{\mu} \rightarrow \nu_{\tau}$
mixing for 
\mbox{$\Delta m^2 = 3 \times 10^{-3}$ eV$^2$}
for a representative 
neutrino energy, $E = 3$ GeV. 
The dashed curve
incorporates energy averaging 
and muon angular smearing
(with respect to the neutrino direction)
and clearly 
fits the data.

More generally,
the (locally averaged)
survival probability
$< \hspace{0.2mm} P_{\mu} \hspace{0.2mm} >$ 
for $\nu_{\mu}$
at intermediate $L/E$ scales,
$(\Delta m'^2)^{-1}$ $\simlt$ $L/E$ $\simlt$ $(\Delta m^2)^{-1}$
(where $L$ is the neutrino flight path length)  
is given by the 
magnitude-squared $|U_{\mu 3}|^2$ 
of the MNS matrix element $U_{\mu 3}$
via
$< \hspace{0.2mm} P_{\mu} 
             \hspace{0.2mm}>$ $ = (1-|U_{\mu3}|^2)^2+|U_{\mu3}|^4$,
whereby
 $|U_{\mu 3}|^2 = [1 \pm \sqrt{1-2(1- \hspace{0.3mm}
       < \hspace{0.2mm} P_{\mu} \hspace{0.2mm} > \hspace{0.3mm} )}]/2$
(clearly $< \hspace{0.2mm} P_{\mu} 
               \hspace{0.2mm} > \hspace{0.3mm} = 1/2$ 
implies
$|U_{\mu 3}|^2 = 1/2$, as expected).
The overall fit to the SK atmospheric data 
(ie.\ not just $<\hspace{0.2mm} P_{\mu} 
                \hspace{0.2mm} >$ $= (U/D)_{\mu}$ 
from the multi-GeV muons above)
gives $< \hspace{0.2mm} P_{\mu} 
                \hspace{0.2mm} >$ $= 0.50 \pm 0.04$
whereby
$0.36$ $\simlt$ $|U_{\mu 3}|^2$ $\simlt$ $0.64$ (68\% CL),
certainly consistent with $|U_{\mu 3}|^2 =1/2$
and twofold maximal mixing. 
Independent evidence for 
strong $\nu_{\tau}$ mixing
comes from the observation
of a substantive
charged-current $\nu_{\tau}$ appearance signal,
statistically separated
from the neutral-current sample
in SUPER-K \cite{tau}. 

As indicated in Section~1,
intermediate baseline reactor experiments,
such as CHOOZ \cite{chooz} and PALO-VERDE \cite{palov},
in fact provide the best limits
on $\nu_e$ (actually $\bar{\nu}_e$) 
mixing at the atmospheric scale
(in terms of the vacuum mixing matrix,
the interpretation of the atmospheric experiments themselves
can be seriously obscured by terrestrial matter effects,
which tend to suppress $\nu_e$-mixing 
and enhance $\nu_{\tau}$-mixing
\cite{hps4}
in the high-energy limit).
Reactor experiments,
utilising very low energy (anti) neutrinos
and with existing baselines much shorter than 
the matter-oscillation length in the Earth,
turn out to be almost
completely immune to matter effects \cite{wgs0}.
As discussed in Section~1,
the best reactor limits
give $|U_{e3}|^2$ $\simlt$ 0.03 (95\% CL) 
for $\Delta m^2 \simeq 3 \times 10^{-3}$~eV$^2$,
consistent with $U_{e3}=0$
and thus with two-fold maximal 
$\nu_{\mu}-\nu_{\tau}$ mixing.

We emphasise that
the atmospheric and reactor 
data do not {\em require}
$U_{e3} \equiv 0$ any more than they
require $|U_{\mu 3}|^2 \equiv 1/2$
(small non-zero values of $U_{\mu 3}$, 
and/or somewhat different 
values of $|U_{\mu 3}|^2$,
eg.\ $|U_{\mu 3}|^2 =2/3$ \cite{fxng}, 
are more-or-less
equally acceptable experimentally).
It is only
that $U_{e3}=0$ and $|U_{\mu3}|^2=1/2$
provide a simple and adequate 
description of the 
current trend of the data, 
making
twofold maximal 
$\nu_{\mu}-\nu_{\tau}$ mixing  
(for now) the accepted
`default option' \cite{jell}
(at the atmospheric scale).

In a similar spirit, 
we turn again to
the solar data displayed in Fig.~1,
drawing particular attention 
now to the solid curve representing 
the `$5/9-1/3-5/9$' bathtub
(in the LMA solution the base of the bathtub
essentially measures $|U_{e2}|^2$ directly).
In Fig.~1 the data are plotted
assuming BP2001 fluxes \cite{bp2001}.
The SNO , HOMESTAKE and SUPER-K data 
(after correction for the neutral-current
contribution in \mbox{SUPER-K})
then determine the base of the bathtub,
with $|U_{e2}|^2 \sim 1/3$ clearly 
closely preferred.
In the flux-independent approach \cite{fog2}
the $^8$B suppression is found to be
\mbox{$\sim  0.33 \pm$
\raisebox{0.9ex}{0.10} \raisebox{-0.9ex}{\hspace{-8.8mm}0.07}}
(based on the measured SK$-$SNO difference \cite{sno}).
In Fig.~1 
the two broken curves correspond
roughly (in the bathtub region)
to the $\pm 1 \sigma$ errors
on the flux-independent supression.
Finally, note that
with $|U_{e3}|^2 = 0$ (or small)
the $\nu_e$
survival probability outside the bathtub, 
$< \hspace{0.1mm} P_e \hspace{0.1mm} >$,
is (inversely) correlated 
with the value at its base 
in the LMA solution.
For vacuum mixing
\mbox{$< \hspace{0.1mm} P_e \hspace{0.1mm} > 
               =(1-|U_{e2}|^2)^2+|U_{e2}|^4$},
so that 
for $|U_{e3}|^2=1/3$ we have
$< \hspace{0.1mm} P_e \hspace{0.1mm} > = 5/9 \simeq 0.56$.
(Taking account of the 
$^8$B contribution the expected gallium 
suppression is actually closer to $\simeq 0.53$,
but this is still consistent with the data
at the $\sim 1.2\, \sigma$ level).
Thus the  gallium data themselves
provide an independent cross-check
on the consistency of the LMA solution
and on $|U_{e2}|^2 \sim 1/3$.

As above, we emphasise that the data
do not {\em require} $|U_{e2}|^2 \equiv 1/3$.
If the (implicit) energy dependence 
of the solar suppression is real,
certainly $|U_{e2}|^2 \ne 1/2$ \cite{jell},
since there can be no MSW bathtub 
in that case \cite{hps2}.
But a somewhat 
more pronounced
(eg. a `$5/8-1/4-5/8$') bathtub 
(corresponding to $|U_{e2}|^2 =1/4$),
is clearly far from excluded.
As before,
we regard $|U_{e2}|^2=1/3$ as 
a simple and adequate
description of the data,
which could usefully come to be seen 
as the default option
at the solar scale. \\

\noindent {\bf 3. `Tri-Bimaximal' Mixing} \\

\noindent
In this section we simply take
the above `default' values $U_{e3}=0$,
$|U_{\mu 3}|^2 = 1/2$
and $|U_{e2}|^2 = 1/3$ as given,
and use them to evaluate the resulting
lepton mixing matrix.

The lepton mixing matrix
is defined
with the rows labelled
by the charged-lepton mass-eigenstates
($e$, $\mu$, $\tau$)
and the columns labelled by the
neutrino mass-eigenstates
($\nu_1$, $\nu_2$, $\nu_3$).
Focussing on the last column
(the $\nu_3$ column),
we note that with $U_{e3} =0$ 
and $|U_{\mu 3}|^2 = 1/2$,
we have $|U_{\tau 3}|^2 = 1/2$ from unitarity,
so that the last column
is just as in
the original bimaximal 
scheme \cite{bmax}.
Moving to the center column
(the $\nu_2$ column),
again as a consequence of $U_{e3}=0$,
orthogonality requires
$|U_{\tau 2}| = |U_{\mu 2}|$.
With $|U_{e2}|^2 =1/3$ (above)
we then have 
$|U_{\tau 2}|^2 = |U_{\mu 2}|^2 =1/3$
from unitarity,
so that the center column
is just as in the
original trimaximal scheme (see eg.\ Ref.~\cite{hps4}).
Finally 
the first column
(the $\nu_1$ column)
follows 
from unitarity
applied to the rows.
 
Indeed, it was 
pointed out in Ref.~\cite{hps4}
(even before the SNO data first appeared)
that a mixing scheme
with the $\nu_3$ `bimaximally' mixed
and the $\nu_2$ `trimaximally' mixed
(hence tri-bimaximal mixing)
could naturally account for the data,
being also discussed
in the conference literature
under the name of
`optimised' bimaximal mixing
\cite{wgs1}:
\begin{eqnarray}
     \matrix{  \hspace{0.1cm} \nu_1 \hspace{0.2cm}
               & \hspace{0.4cm} \nu_2 \hspace{0.2cm}
               & \hspace{0.4cm} \nu_3  \hspace{0.2cm} }
                                      \hspace{0.4cm} \nonumber \\
(|U_{l \nu}|^2) \hspace{2mm} = \hspace{2mm}
\matrix{ e \hspace{0.2cm} \cr
         \mu \hspace{0.2cm} \cr
         \tau \hspace{0.2cm} }
\left( \matrix{ 2/3  &
                      1/3 &
                              0 \cr
                1/6 &
                    1/3 &
                             1/2  \cr
      \hspace{2mm} 1/6 \hspace{2mm} &
         \hspace{2mm}  1/3 \hspace{2mm} &
           \hspace{2mm} 1/2  \hspace{2mm} \cr } \right)
\label{obx}
\end{eqnarray}
(where the moduli-squared of the elements are given).
The name
`optimised' bimaximal mixing
reflected the scheme's pedigree
as a special case of the
Altarelli-Feruglo generalised bimaximal form \cite{altf}
and its close relationship
to the original bimaximal form \cite{bmax}.
We emphasise that the 
mixing Eq.~1
is entirely determined
by unitarity constraints once the above three
`corner' elements are
fixed to their default values.

We should also point out
that the mixing Eq.~1 has much in common
with the Fritzsch-Xing democratic ansatz 
\cite{fxng}
(which might in fact, see Section~4, 
reasonably be termed
`bi-trimaximal' mixing,
as opposed to 
`tri-bimaximal' mixing).
Indeed 
the Fritzsch-Xing ansatz
may be viewed as
a permuted form
of Eq.~1
(with, somewhat remarkably,
the 
crucial prediction $U_{e3} = 0$
made well before 
the emergence of the CHOOZ data \cite{chooz}).
It should be clear however that
the phenomenologies of these 
two mixing shemes
are quite distinct,
eg. the Fritzsch-Xing ansatz
predicted $|U_{e2}|^2=1/2$ 
and hence no energy dependence \cite{hps2}
of the solar suppression,
which is now disfavoured 
by SNO \cite{sno}.
The original 
bimaximal scheme \cite{bmax}
is likewise now disfavoured \cite{bah1}.

Asymptotic 
($L/E \rightarrow \infty$) predictions,
specific to tri-bimaximal mixing,
are the (vacuum) survival probabilites 
$< \hspace{0.1mm} P_{\mu} \hspace{0.1mm} > 
\hspace{1.5mm} = \hspace{1.5mm}
 < \hspace{0.1mm} P_{\tau} \hspace{0.1mm} > 
\hspace{1.5mm} = \hspace{1.5mm} 7/18$.
The corresponding $\nu_{\mu} \leftrightarrow \nu_{\tau}$ 
appearance probability is also $7/18$.
Note that $U_{e3} \equiv 0$
implies no Pantaleone resonance \cite{pant}
and no CP violation
in neutrino oscillations,
which might be
considered 
a disappointment
experimentally.
Nonetheless,
it is fair to say
that current data
{\em point} to Eq.~1,
and it is therefore
of interest 
to try to understand
what it might imply.
In the next section
we present 
simple mass-matrices 
leading to tri-bimaximal mixing. \\

\noindent {\bf 4.  
Simple Mass Matrices } \\

\noindent 
Our mass-matrices
will be taken to be hermitian
(ie. we will throughout 
be implicitly referring 
to hermitian-squares 
of mass-matrices
linking left-handed fields,
$MM^{\dagger} \equiv M^2$). 
Fermion mass-matrices
are most naturally considered
in a `weak' basis
(i.e. a basis which leaves 
the charged-current
weak-interaction 
diagonal and universal).

Maximal mixing 
(whether trimaximal or bimaximal)
undeniably
suggests permutation symmetries \cite{hs1}.
We postulate that,
in a particular weak basis,
the mass-matrices 
take the following 
(permuation symmetric) 
forms:
\begin{equation}
M_l^2 =
\left(\matrix{
a & b & b^{*} \cr
b^{*} & a & b \cr
b & b^{*} & a \cr
} \right)
\hspace{2.0cm}
M_{\nu}^2 =
\left(\matrix{
x & 0 & y \cr
0 & z & 0 \cr
y & 0 & x \cr
} \right) \hspace{2.0cm} 
\label{m2}
\end{equation}
where the real constants $a$, $x$, $y$, $z$
and the complex constants $b$ and $b^*$ 
encode
the charged-lepton and neutrino masses as follows:
\begin{eqnarray}
a=\frac{m_e^2}{3} +\frac{m_{\mu}^2}{3}+\frac{m_{\tau}^2}{3} 
                     \hspace{2.9cm}
     x=\frac{m_1^2}{2} +\frac{m_{3}^2}{2} \hspace{1.8cm} \nonumber \\
b=\frac{m_e^2}{3} +\frac{m_{\mu}^2 \omega}{3}
                  +\frac{m_{\tau}^2 \bar{\omega}}{3} 
                                 \hspace{2.9cm}
     y=\frac{m_1^2}{2} -\frac{m_{3}^2}{2}  \hspace{1.8cm} \label{abc} \\
b^*=\frac{m_e^2}{3} +\frac{m_{\mu}^2 \bar{\omega}}{3}
                  +\frac{m_{\tau}^2 \omega}{3} 
                                      \hspace{2.9cm}
     z=m_2^2 \hspace{1.2cm} \hspace{1.8cm} \nonumber
\end{eqnarray}
with $\omega=\exp(i2\pi/3)$ and $\bar{\omega}=\exp(-i2 \pi/3)$
denoting the complex cube roots of unity.

In Eq.~\ref{m2}
the charged-lepton mass-matrix $M_l^2$
takes the familiar $3 \times 3$ circulant form \cite{hs1},
invariant under cyclic permutations
of the three generation indices.
The neutrino mass-matrix $M_{\nu}^2$ is real 
(i.e.\ symmetric, since our mass-matrices are hermitian)
and is a $2 \times 2$ circulant
in the $1-3$ index subset,
invariant under the permutation 
of only {\em two} 
out of the three generation indices
(generations $1 \leftrightarrow 3$).
The neutrino mass-matrix 
has four `texture zeroes' \cite{ross}  
enforcing the
effective block-diagonal form.
Note that {\em both}
 mass matrices (Eq.~\ref{m2})
are invariant under the
interchange of generation indices $1 \leftrightarrow 3$
performed simultaneously with a complex conjugation.
Indeed, in this basis,
it is the invariance
of {\em all} the leptonic terms
under this combined involution,
which guarantees no CP-violation here 
(since Im $\det [M_l^2,M_{\nu}^2]$
\cite{jarl} changes sign).

The mass-matrices $M_l^2$ and $M_{\nu}^2$
are diagonalised 
by a threefold maximal unitary matrix $U_l$
and a twofold maximal unitary matrix $U_{\nu}$,
respectively:
\begin{eqnarray}
    \matrix{  \hspace{0.2cm} e \hspace{0.2cm}
               & \hspace{0.2cm} \mu \hspace{0.1cm}
               & \hspace{0.2cm} \tau  \hspace{0.2cm} }
                                      \hspace{0.9cm}
\hspace{1.9cm}
    \matrix{  \hspace{0.2cm} \nu_1 \hspace{0.1cm}
               & \hspace{0.2cm} \nu_2 \hspace{0.1cm}
               & \hspace{0.2cm} \nu_3  \hspace{0.2cm} }
                                      \hspace{0.4cm} \nonumber \\
U_l=
\left(\matrix{
\frac{1}{\sqrt{3}}
                & \frac{\bar{\omega}}{\sqrt{3}} 
                           & \frac{\omega}{\sqrt{3}} \cr
\frac{1}{\sqrt{3}}
                & \frac{1}{\sqrt{3}}
                    & \frac{1}{\sqrt{3}} \cr
\frac{1}{\sqrt{3}}
             & \frac{\omega}{\sqrt{3}}
                    & \frac{\bar{\omega}}{\sqrt{3}} \cr
} \right)
\hspace{1.5cm}
U_{\nu}=
\left(\matrix{
\sqrt{\frac{1}{2}} & 0 & -\sqrt{\frac{1}{2}} \vspace{5pt} \cr
 0 & 1 & 0 \vspace{4pt} \cr
\sqrt{\frac{1}{2}} & 0  & \sqrt{\frac{1}{2}} \cr
} \right)
\hspace{0.35cm}
\label{ulun}
\end{eqnarray}
i.e.\ $U_l^{\dagger}M_l^2U_l$ $=$ 
diag ($m_e^2$, $m_{\mu}^2$, $m_{\tau}^2$) and
$U_{\nu}^{\dagger}M_{\nu}^2U_{\nu}$ $=$ 
diag ($m_1^2$, $m_2^2$, $m_3^2$),
so that
the lepton mixing matrix
(or MNS matrix)
$U=U_l^{\dagger}U_{\nu}$ is given by:
\begin{eqnarray}
    \matrix{  \hspace{0.2cm} \nu_1 \hspace{0.1cm}
               & \hspace{0.2cm} \nu_2 \hspace{0.1cm}
               & \hspace{0.2cm} \nu_3  \hspace{0.2cm} }
                                      \hspace{0.9cm}
\hspace{1.5cm}
    \matrix{  \hspace{0.2cm} \nu_1 \hspace{0.1cm}
               & \hspace{0.2cm} \nu_2 \hspace{0.2cm}
               & \hspace{0.2cm} \nu_3  \hspace{0.2cm} }
                                      \hspace{1.0cm} \nonumber \\
\matrix{ e \hspace{0.2cm} \cr
         \mu \hspace{0.2cm} \cr
         \tau \hspace{0.2cm} }
\left(\matrix{
\frac{1}{\sqrt{3}}
                & \sqrt{\frac{1}{3}} & \frac{1}{\sqrt{3}} \cr
\frac{\omega}{\sqrt{3}}
                & \sqrt{\frac{1}{3}}
                    & \frac{\bar{\omega}}{\sqrt{3}} \cr
\frac{\bar{\omega}}{\sqrt{3}}
             & \sqrt{\frac{1}{3}}
                    & \frac{\omega}{\sqrt{3}} \cr
} \right)
\hspace{0.2cm}
\left(\matrix{
\sqrt{\frac{1}{2}} & 0 & -\sqrt{\frac{1}{2}} \vspace{5pt} \cr
0 & 1 & 0 \vspace{4pt} \cr
\sqrt{\frac{1}{2}} & 0  & \sqrt{\frac{1}{2}} \cr
} \right)
\hspace{0.35cm}
=
\hspace{0.35cm}
\matrix{ e \hspace{0.2cm} \cr
         \mu \hspace{0.2cm} \cr
         \tau \hspace{0.2cm} }
\left(\matrix{
\sqrt{\frac{2}{3}} & \sqrt{\frac{1}{3}} &  0 \cr
-\sqrt{\frac{1}{6}} & \sqrt{\frac{1}{3}} & -\frac{i}{\sqrt{2}} \cr
-\sqrt{\frac{1}{6}}& \sqrt{\frac{1}{3}}   & \frac{i}{\sqrt{2}} \cr
} \right) \hspace{0.80cm}
\label{decomp}
\end{eqnarray}
where the RHS is
the tri-bimaximal form (Eq.~\ref{obx})
in a particular phase convention.
For Dirac neutrinos,
the factor of $i$ is readily
removed by a simple rephasing
of the $\nu_3$ mass-eigenstate,
yielding tri-bimaximal mixing
expressed as an orthogonal matrix:
\begin{eqnarray}
     \matrix{  \hspace{0.4cm} \nu_1 \hspace{0.2cm}
               & \hspace{0.2cm} \nu_2 \hspace{0.2cm}
               & \hspace{0.2cm} \nu_3  \hspace{0.4cm} }
                                      \hspace{1.4cm} \nonumber \\
\hspace{1.0cm}U \hspace{0.3cm} = \hspace{0.3cm} 
\matrix{ e \hspace{0.2cm} \cr
         \mu \hspace{0.2cm} \cr
         \tau \hspace{0.2cm} }
\left( \matrix{ \sqrt{\frac{2}{3}}  &
                      \frac{1}{\sqrt{3}} &
                                   0  \cr
 - \frac{1}{\sqrt{6}}  &
          \frac{1}{\sqrt{3}} &
  \frac{1}{\sqrt{2}}   \cr
      \hspace{2mm} 
  - \frac{1}{\sqrt{6}} \hspace{2mm} &
         \hspace{2mm} 
            \frac{1}{\sqrt{3}} \hspace{2mm} &
                                 -\frac{1}{\sqrt{2}}  
                                       \hspace{2mm} \cr } \right).
\hspace{1.0cm}
\label{tbx}
\end{eqnarray}

While this concludes our 
derivation of tri-bimaximal mixing
starting from Eq.~\ref{m2},
we take this opportunity 
to remark that one might easily
(with perhaps equal {\it a priori} justification)
have interchanged the forms
of $M_l^2$ and $M_{\nu}^2$
in Eq.~\ref{m2}, 
taking the {\em neutrino} mass-matrix
to be the $3 \times 3$ circulant,
and the charged-lepton
mass-matrix to be of the 
$2 \times 2$ block-diagonal 
circulant form.
Note however
that this leads to physically distinct 
mixing
which, if the $2\times 2$ circulant
is chosen to be in the $1-2$ index subset,
is identically the Fritzsch-Xing 
democratic ansatz \cite{fxng}
(hence `bi-trimaximal' mixing
as a synonym for the Fritzsch-Xing ansatz,
see Section~3 above). 
We emphasise once again that
the phenomenologies of 
the above two mixing schemes
are physically distinct,
with the Fritzsch-Xing ansatz
now essentially ruled out, 
along with many other schemes involving
energy independent solar solutions \cite{roy}
(including trimaximal mixing \cite{hps1}) 
following the SNO results \cite{sno}. \\

\noindent {\bf 5. Perspective } \\

\noindent
Tri-bimaximal mixing 
is a specific mixing matrix 
(Eq.~\ref{obx}/Eq.~\ref{tbx}) 
which encapsulates
the trends of a broad range
of experimental data
(the LSND 
oscillation
signal \cite{lsnd}
was not considered
on the grounds that
it still awaits 
confirmation).
Tri-bimaximal mixing is closely
related to a number of previously
suggested 
lepton mixing schemes,
notably trimaximal mixing \cite{hps1},
bimaximal mixing \cite{bmax},
the Fritzsch-Xing democratic ansatz \cite{fxng}
and the Altarelli-Feruglio scheme \cite{altf}
(of which tri-bimaximal mixing
may be considered 
a special case).
For the future,
the couplings of the heavy neutrino, $\nu_3$,
are expected to be measured more precisely
in long-baseline experiments like MINOS \cite{minos}
(and other projects \cite{blond}).
In particular, the limits
on $|U_{e3}|^2$
should continue to improve.
Regarding the $\nu_2$ couplings,
if tri-bimaximal mixing is right,
the KAMLAND experiment \cite{kaml} should
confirm the LMA solution,
measuring a $\bar{\nu}_e$ survival probability 
tending to $< \hspace{0.0mm} P_e \hspace{0.0mm} > 
= 5/9 \simeq 0.56$ 
at sufficiently low-energy.
Corresponding predictions
for $\nu_{\mu}$ disappearance
and $\nu_{\mu} \rightarrow \nu_{\tau}$
appearance, see Section 3,
look very hard to test
(experiments with $\nu_{\mu}$
necessitate higher energies
and hence longer baselines,
with matter effects 
generally dominant over vacuum effects
for $\nu_2$).
Finally, {\em exact}
tri-bimaximal mixing
would imply
no high-energy 
matter resonance
and no 
(intrinsic) CP-violation
in neutrino oscillations,
which might be 
considered a disappointment experimentally. \\

\vspace{5mm}
\noindent {\bf Acknowledgement}

\noindent It is a pleasure to thank
 A.\ Astbury for a helpful suggestion.
This work was supported by the
UK Particle Physics and Astronomy
Research Council (PPARC).

\newpage

\noindent {\bf {\large Figure Captions}}

\vspace{10mm}
\noindent Figure~1.
The solar data, 
including the recent SNO measurements,
plotted versus neutrino energy $E$.
The gallium points 
are plotted at $<1/E>^{-1} \simeq 0.5$ MeV,
the SUPER-K point at $<1/E>^{-1} \simeq 10$ MeV
and the HOMESTAKE point
(which includes an $\sim 15\%$
contribution from the $^7$Be-line
at $\sim 0.9$ MeV) is plotted 
(somewhat arbitrarily)
at $<1/E>^{-1} = 5$ MeV.
The solid curve is
the expected energy dependence
for tri-bimaximal mixing,
Eq.~1, with
$\Delta m'^2 = 5 \times 10^{-5}$ eV$^2$,
which predicts
a suppression of $1/3$
in the base of the bathtub
and $5/9$ outside.

\vspace{10mm}
\noindent Figure~2.
Atmospheric neutrino data.
The multi-GeV zenith angle distributions
for a) $\mu$-like and b) $e$-like events
in SUPER-K.
The solid curve is the full oscillation curve
for tri-bimaximal mixing (Eq.~1/Eq.~6)
with $\Delta m^2 = 3 \times 10^{-3}$ eV$^2$
for a representative neutrino energy $E = 3$ GeV,
and the dashed curve 
shows the effect of angular smearing
with respect to the neutrino direction
and averaging over neutrino energies.
Tri-bimaximal mixing
(or indeed any mixing hypothesis
which is
effectively twofold maximal
$\nu_{\mu}-\nu_{\tau}$ mixing
at the atmospheric scale)
fits the data well.

\newpage
\pagestyle{empty}
\begin{figure*}[hbt]
\epsfig{figure=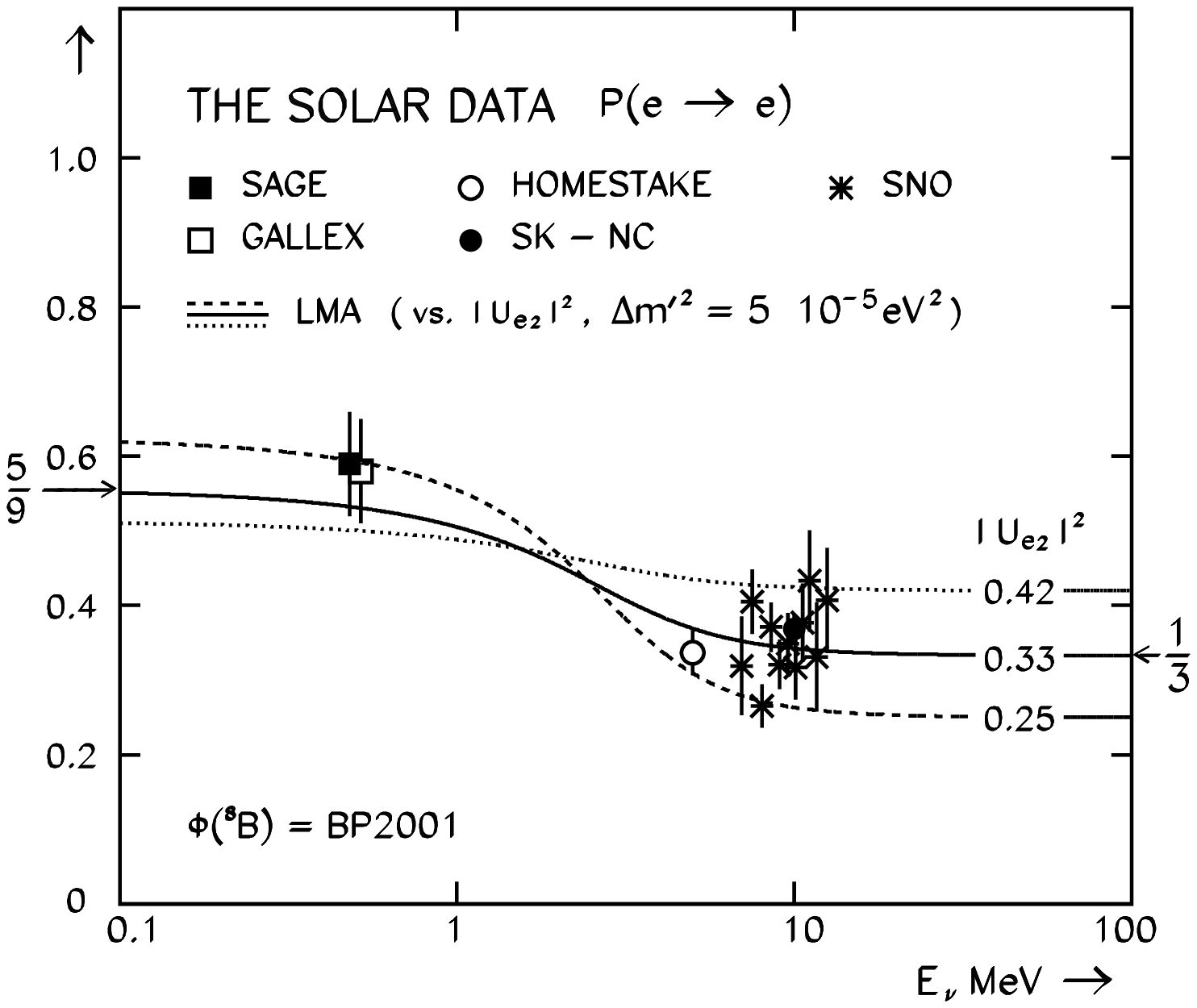
,width=150mm,bbllx=80pt,bblly=120pt
,bburx=530pt,bbury=720pt}
\caption{ }
\end{figure*}


\newpage
\pagestyle{empty}
\begin{figure*}[hbt]
\epsfig{figure=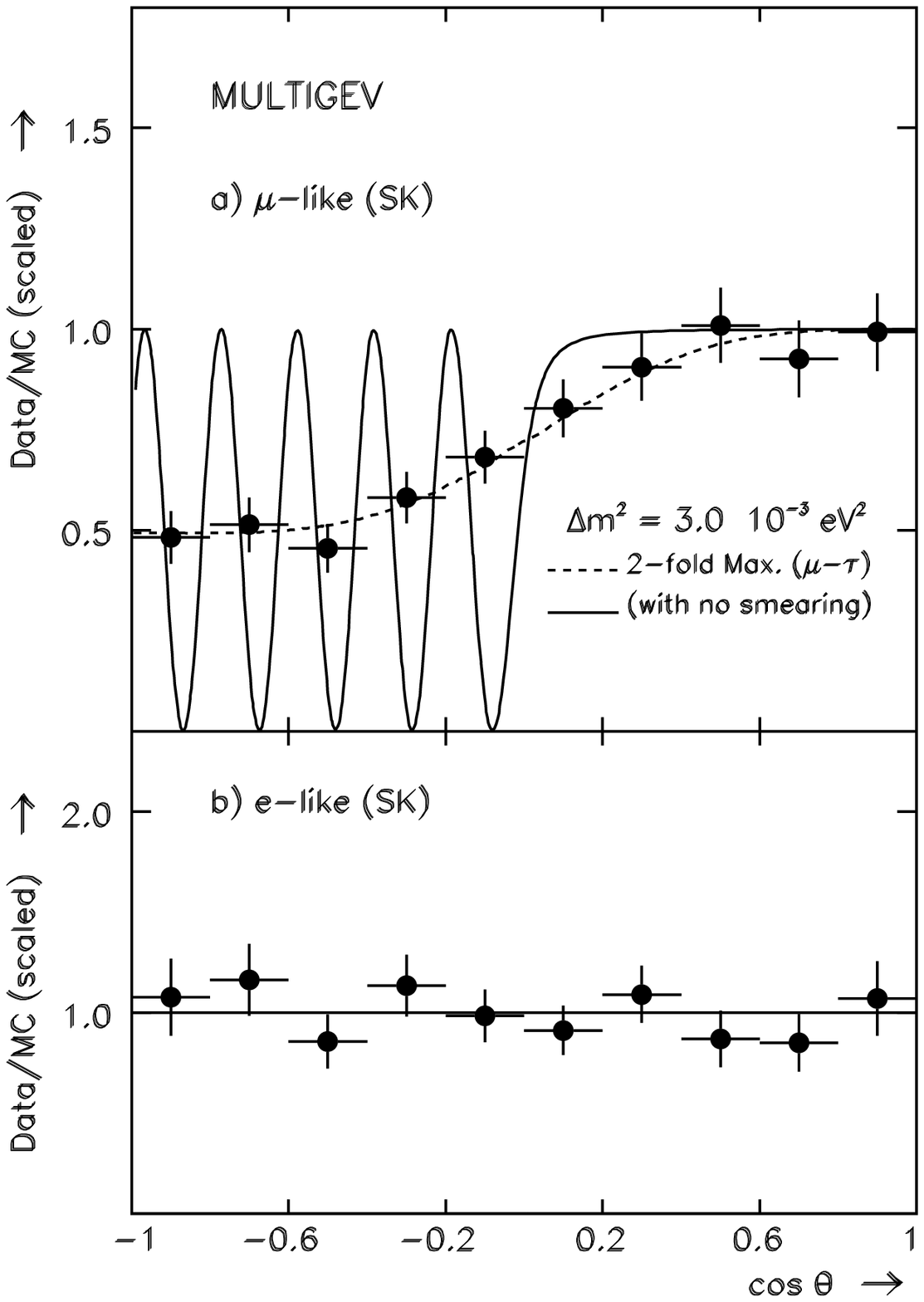
,width=150mm,bbllx=80pt,bblly=120pt
,bburx=530pt,bbury=720pt}
\caption{ }
\end{figure*}


\newpage

\begin{thebibliography}{99}
\bibitem[1]{sno} Q.\ R.\ Ahmad et al.\ 
                 Phys.\ Rev.\ Lett.\ 87 (2001) 071301
                 (nucl-ex/0106015).
\bibitem[2]{home} B.\ T.\ Cleveland et al.\
                  AstroPart.\ Phys.\ 496 (1998) 505.
\bibitem[3]{sk1258}  S.\ Fukuda  et al.\
                     Phys.Rev.Lett. 86 (2001) 5651
                     (hep-ex/01030302). \\
                     S.\ Fukuda et al.\ 
                     Phys.Rev.Lett. 86 (2001) 5656
                     (hep-ex/0103033)
\bibitem[4]{bp2001}  J.\ N.\ Bahcall, M.\ H. Pinsonneault and S.\ Basu.
                     Astropys.\ J.\ 555 (2001) 990 (astro-ph/0010346).
\bibitem[5]{fog1}  G.\ L.\ Fogli  et al.\ Phys Rev D 64 (2001) 0093007
                   (hep-ph/0106247).
\bibitem[6]{sage} J. Abdurashitov et al.
                  Phys Rev C60 (1999) 055801
\bibitem[7]{gall} J. W. Hampel et al.
                  Phys. Lett. B 447 (1999) 127.
\bibitem[8]{gno} M. Altmann et al.
                  Phys. Lett. B 490 (2000) 16 (hep-ex/0006034).
\bibitem[9]{hps1} P.\ F.\ Harrison, D.\ H.\ Perkins and W.\ G.\ Scott.
                  Phys.\ Lett.\ B 349 (1995) 357.
\bibitem[10]{msw} L. Wolfenstein, Phys. Rev. D17 (1978) 2369;
                                  D20 (1979) 2634. \\
                 S. P. Mikheyev and A. Yu. Smirnov,
                 Il Nuovo Cimento 9C (1986) 17.
\bibitem[11]{hps2} P.\ F.\ Harrison, D.\ H.\ Perkins and W.\ G.\ Scott.
                  Phys.\ Lett.\ B 374 (1996) 111 (hep-ph/9601346).
\bibitem[12]{atm} Y.\ Fukuda et al.\ Phys.\ Let.\ B 436 (1998) 33
                  (hep-ex/9805006). \\
                  W.\ W.\ M.\ Allison et al.\ Phys.\ Lett. B (1999) 137
                  (hep-ex/9901024).
\bibitem[13]{k2k} S.\ H.\ Ahm et al.\ Phys.\ Lett.\ B 511 (2001) 178
                  (hep-ex/0103001).
\bibitem[14]{fog2} G.\ L.\ Fogli, E.\ Lisi and A. Marrone.
                   hep-ph/0110089.  
\bibitem[15]{chooz} M. Apollonio et al.\ Phys.\ Lett.\ B 420 (1998) 397
                   (hep-ex/9711002); \\
                    Phys.\ Lett.\ B.\ 466 (1999) 415 (hep-ex/9907037).
\bibitem[16]{palov} F.\ Boehm et al.\ Phys.\ Rev.\ D64 (2001) 112001
                    (hep-ex/0107009).
\bibitem[17]{mns}  Z.\ Maki, M.\ Nakagawa and s.\ Sakata
                   Prog.\ Theor.\ Phys.\ 28 (1962) 870. 
\bibitem[18]{tau}  S.\ Fukuda et al.\ Phys.\ Rev.\ Lett.\ 85 (2000) 3999 
                   (hep-ex/0009001).
\bibitem[19]{hps4} P.\ F.\ Harrison, D.\ H.\ Perkins and W.\ G.\ Scott.
                   Phys.\ Lett.\ B 458 (1999) 79. (hep-ph/9904297).
\bibitem[20]{wgs0}  W.\ G.\ Scott. Proc.\ 2nd Int.\ Work.\
                    Ident.\ Dark Matter (IDM98) Buxton, UK. \\
                    ed.\ N.\ Spooner, V.\ Kudryatsev.
                    World Scientific (1999) 540. RAL-TR-1998-072.
\bibitem[21]{fxng} H.\ Fritzsch and Z.\ Xing 
                   Phys.\ Lett.\ B 372 (1996) 265 (hep-ph/9509389); \\
                   Phys.\ Lett.\ B 440 (1999) 313 (hep-ph/9808272); \\
                   Phys.\ Rev.\ D61 (2000) 073016 (hep-ph/9909304).
\bibitem[22]{jell}  J.\ Ellis `Neutrino 2000'
                    Nucl.\ Phys.\ (Proc.\ Suppl.\ )
                    B91 (2000) 503 \\ (hep-ph/0008334).
\bibitem[23]{bmax}   F.\ Vissani. hep-ph/9708483 (unpublished). \\
                    V.\ Barger et al.
                    Phys.\ Lett.\ B437 (1998) 107
                    (hep-ph/9806387). \\
                    A.\ J.\ Bahz, A.\ S.\ Goldhaber and M.\ Goldhaber. \\
                    Phys.\ Rev.\ Lett.\ 81 (1998) 5730 (hep-ph/9806540). \\
                    D.\ V.\ Ahluwalla. Mod.\ Phys.\ Lett.\ A 18 (1998) 2249. \\
                    H.\ Giorgi and S.\ L.\ Glashow. hep-ph/9808293.
\bibitem[24]{wgs1}  W.\ G.\ Scott. 6th Topic.\ Sem.\ 
                    Neutrinos AstroPart.\ Phys. 
                    San Miniato, Italy. \\
                    Nucl.\ Phys.\ B (Proc.\ Suppl.\ ) 85 (2000) 177
                    (hep-ph/9909431). \\  
                    W.\ G.\ Scott. Proc.\ 3rd Int.\ Work.\ 
                    Ident.\ Dark Matter (IDM2000) York, UK. \\
                    ed.\ N.\ Spooner, V.\ Kudryatsev.
                    World Scient.\ (2001) 526 (hep-ph/0010335).
\bibitem[25]{altf}  G.\ Altarelli and F.\ Feruglio
                    Phys.\ Lett.\ B439 (1998) 112 (hep-ph/9807353); \\ 
                    Phys.\ Lett.\ B451 (1999) 388 (hep-ph/9912475).
\bibitem[26]{bah1}  J.\ N.\ Bahcall, M.\ C.\ Gonzalez-Garcia
                    and C.\ Pena-Garay. hep-ph/0111150. 
\bibitem[27]{pant}  J.\ Pantaleone.
                    Phys.\ Rev.\ Lett.\ 81 (1998) 5060 (hep-ph/9810467). 
\bibitem[28]{hs1}   P.\ F.\ Harrison and W.\ G.\ Scott
                    Phys.\ Lett.\ B 333 (1994) 471 (hep-ph/9406351). \\
                    E.\ Derman and D.\ R.\ T.\ Jones 
                    Phys.\ Lett.\ B70 (1977) 449. \\
                    C.\ Lee, C.\ Lin and Y. Yang. 
                    Phys.\ Rev.\ D42 (1990) 2355. \\
                    S.\ L.\ Adler Phys. Rev. D59 (1999) 015012 
                    (hep-ph/9806518).
\bibitem[29]{ross}  P.\ Ramond, R.\ G.\ Roberts and G.\ G.\ Ross.
                    Nucl.\ Phys.\ B 406 (1993) 19 \\ (hep-ph/9303320). 
\bibitem[30]{jarl}  C.\ Jarlskog Z.\ Phys.\ C 29 (1985) 491;
                    Phys.\ Rev.\ Lett.\ 55 (1985) 1039.
\bibitem[31]{roy}   S.\ Choubey et al.\ Phys.\ Rev.\ D64 (2001) 
                    052002 (hep-ph/0103318); \\
                    S.\ Choubey et al.\ hep-ph/0109017. \\
                    A.\ Bandyopadhyay et al.\ hep-ph/0110307.
\bibitem[32]{lsnd}  C.\ Athanassopoulos et al.
                    Phys.\ Rev.\ Lett.\ 75 (1995) 2650 (hep-ex/9504002); \\ 
                    Phys.\ Rev.\ Lett.\
                    77 (1996) 3082 (hep-ex/9605003); \\
                    Phys.\ Rev.\ Lett.\
                    81 (1998) 1774 (hep-ex/9709006).
\bibitem[33]{minos} E.\ Ables et al.\ (MINOS collab.)
                    Fermilab-Propoasal P-875 (1995).
\bibitem[34]{blond} J.\ J.\ Gomez Cadenas et al.\ 
                    9th Intl. Symposium on Neutrino Telescopes. \\ 
                    Venice (2001) (hep-ph/0105297). \\
                    A.\ Blondel et al.\
                    Nucl.\ Inst.\ Methods.\ A451 (2000) 102. \\
                    T.\ R.\ Edgecock and W.\ J.\ Murray.
                    J.\ Phys.\ G.\ Nucl.\ Part.\ Phys.\ 27 (2001) R141.
\bibitem[35]{kaml} A.\ Suzuki. 8th Intl.\ Workshop
                   Neutrino Telescopes. Venice (1999).
\end{thebibliography}
\end{document}